\documentclass[a4paper, twocolumn]{article}
\usepackage{bm}
\usepackage{amssymb,amsmath,graphicx,bbm,color,booktabs}
\usepackage{amsopn}

\newcommand{\vx}{\bm{x}}

\renewcommand{\S}{\mathcal{S}}

\renewcommand{\eqref}[1]{Eq.~(\ref{#1})}

\def \x{{\mathbf x}}

\usepackage{INTERSPEECH2022}
\usepackage{dsfont}
\usepackage{color}
\usepackage{MnSymbol} 
\usepackage{graphicx}
\usepackage{multicol}
\usepackage{multirow}
\usepackage{color}
\graphicspath{ {./LaTeX/figures }}

\title{Unsupervised Symbolic Music Segmentation using \\ 
Ensemble Temporal Prediction Errors}
\name{Shahaf Bassan, Yossi Adi, Jeffrey S. Rosenschein}
\address{Hebrew University of Jerusalem}
\email{shahaf.bassan@mail.huji.ac.il}
\begin{document}

\maketitle
\begin{abstract}
Symbolic music segmentation is the process of dividing symbolic melodies into smaller meaningful groups, such as melodic phrases. We proposed an unsupervised method for segmenting symbolic music. The proposed model is based on an ensemble of temporal prediction error models. During training, each model predicts the next token to identify musical phrase changes. While at test time, we perform a peak detection algorithm to select segment candidates. Finally, we aggregate the predictions of each of the models participating in the ensemble to predict the final segmentation. Results suggest the proposed method reaches state-of-the-art performance on the Essen Folksong dataset under the unsupervised setting when considering F-Score and R-value. We additionally provide an ablation study to better assess the contribution of each of the model components to the final results. As expected, the proposed method is inferior to the supervised setting, which leaves room for improvement in future research considering closing the gap between unsupervised and supervised methods.
\end{abstract}

\vspace{-1em}
\section{Introduction}
\label{sec:intro}

Symbolic music segmentation is the task of dividing symbolic melodies into smaller meaningful groups. These can be groups such as the chorus, verse, intro, outro, and bridge in popular music or maybe smaller groups such as phrases or even motifs. Symbolic music segmentation is an important task in Music Information Retrieval (MIR) and it serves as a basis for other applicative tasks, such as melodic feature computation, melody indexing, and retrieval of melodic excerpts~\cite{pearce2010melodic} as well as for the use of musicological research~\cite{mullensiefen2008high}. Measuring the performance of an automatic segmentation method is usually done via its degree of agreement within human annotators. The task of music segmentation is made more difficult by the fact that there is more than a single possible segmentation solution. Certain boundaries can be subjective or ambiguous, a fact demonstrated by multiple researchers who have compared music segmentations of both listeners and musicians~\cite{deliege1987grouping, peretz1989clustering, thom2002melodic, bozkurt2014usul, hartmann2017interaction}.

Previous studies in MIR have suggested different methods to automatically segment musical phrases~\cite{cenkerova2018crossing, bod2002memory, pearce2010melodic}, considering rule-based systems, supervised learning, and unsupervised learning. As annotating music is known to be both time-consuming, expensive~\cite{wang2017re}, and subjective in nature, labeled datasets of segmented symbolic music are hardly found. Hence, there is a growing need for unsupervised segmentation methods for handling this task.

\begin{figure}
	\includegraphics[width=1.0\linewidth]{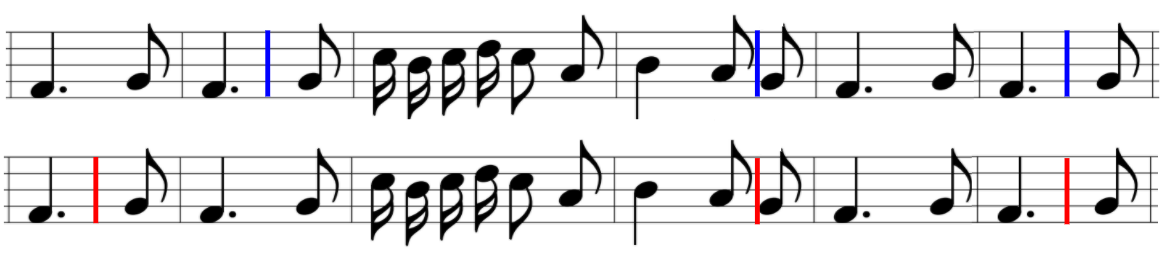}
	\caption{\textit{An example for a segmentation of a song from our test set. The human label segmentation (blue) and our model segmentation results (red))}}
	\label{fig:teaser}
\vspace{-1.8em}
\end{figure}

Segmentation based on temporal prediction errors is the process of detecting peaks in the error curve of a model trained to predict a sequence frame by frame as potential boundaries. This method was first tested and shown to be significant as a method for character-based text segmentation~\cite{elman1990finding, christiansen1998learning}, but recently showed its great potential for phoneme and word speech segmentation~\cite{michel2016blind, wang2017gate, kreuk2020self, cuervo2021contrastive}. Inspired by this work, we explore an unsupervised ensemble method for symbolic music segmentation based on temporal prediction errors. See Figure~\ref{fig:teaser} of visual description of the task.

\noindent {\bf Our Contribution:} (i) We show for the first time the use of temporal prediction error models, and provide an analysis, in music segmentation, specifically for symbolic music phrase segmentation. (ii) We show that the proposed approach reaches state-of-the-art unsupervised results both with relation to F-Score and R-Value on the Essen Folksong Dataset. (iii) We show the potential of using ensemble learning as a method to enhance the performance of temporal prediction error methods for symbolic music segmentation.

\vspace{-0.2em}
\paragraph*{\textbf{Musical Notations}}
\leavevmode
\\
\vspace{-0.5em}
\\
We start by briefly describing some of the used musical notations throughout the paper.
\\
\noindent {\bf Musical phrase} is defined as a unit of musical meter that has a complete musical sense of its own and is considered one of the most important units of music content~\cite{lerdahl1996generative}. 

\noindent {\bf Bar} in music is a segment of time corresponding to a specific number of beats, which is part of the score input.

\noindent {\bf Time signature} (also known as meter signature) is a notational convention used in Western musical notation to specify how many beats or pulses are contained in each musical bar.
\vspace{-0.5em}
\section{Related Work}
\label{sec:related}

Tenney and Polansky~\cite{tenney1980temporal} were perhaps the first to suggest models of symbolic melody segmentation based on musicological rule-based systems. Other common musicological rule-based systems include Grouper by Temperley~\cite{temperley2004cognition}, the preference rules for grouping as defined in A Generative Theory of Tonal Music (GTTM)~\cite{lerdahl1996generative}, the quantization of the GTTM offered by~\cite{frankland2004parsing}, the Implication-Realization
theory by Narmour~\cite{narmour1990analysis, narmour1992analysis} and its quantization~\cite{schellenberg1996expectancy}. More recent rule-based segmentation methods were offered by Lopez~\cite{rodriguez2016automatic} and Cenkerova~\cite{cenkerova2018crossing}.
A common rule-based method that is widely used (also relevant to our approach) is the Local Boundary Detection Model (LBDM)~\cite{cambouropoulos2001local}, which operates on the basis of identifying segment points on local changes in pitch, Inter Onset Intervals (IOI's), and rests. Whereas that model looks at unexpected changes in a piece based on rule-based approaches, it does not employ any capability of identifying these changes based on prediction errors of a machine learning prediction model.

Other approaches that were offered are based on computational or learning algorithms. The Data-Oriented Parsing (DOP)-Markov parser~\cite{bod2002memory} learns probabilities to rewrite rules from a set of examples. It was shown that this method is able to detect phrase-ending patterns that were not shown on musicological rule-based systems. Another work that is relevant to ours is that of Juhasz~\cite{juhasz2004segmentation} that uses a memory-based maximum entropy model for finding less predictable areas in a score that are marked as segments. Other models combined rule-based systems with computational concepts~\cite{ahlback2004melody, cambouropoulos2006musical, ferrand2003memory, rodriguez2016automatic, van2020rule, conklin2006melodic}. More known works include the multiple viewpoint method IDyOM ~\cite{pearce2010melodic}, using a Restricted Boltzman machine to model the probability of melodic events~\cite{lattner2015probabilistic}, and using a CNN or an LSTM together with Conditional Random Field (CRF) for supervised learning of segment points~\cite{zhang2020symbolic, guan2018melodic}.

Recent work shows improvement in unsupervised \emph{speech segmentation} based on the use of temporal prediction errors. Michel et al.~\cite{michel2016blind} showed a model training of a next-frame prediction model using HMM or RNN. Areas that had a high prediction error were chosen as segments using peak detection and tagged as boundaries of phonemes. More recently, Wang et al.~\cite{wang2017gate} trained an RNN autoencoder and tracked the norm of various intermediate gate values (forget-gate for LSTM and update-gate for GRU). To identify boundaries of phonemes, a peak detection technique was used on the gate norm over time. Kreuk et al.~\cite{kreuk2020self} trained a CNN encoder for distinguishing between adjacent frame pairs and random pairs of distractor frames, where a peak detection algorithm is used to detect the phoneme boundaries on the outputs of the model. Our work uses these new concepts, that were shown to be highly effective for phoneme and word segmentation, on music boundary segmentation, and more specifically on music phrase segmentation.
\vspace{-1.25em}
\begin{table} [t!]
    \caption{Essen Folksong distribution over time signatures}
	\label{tab:db_stats}
	\small
	\centering
	\scalebox{0.8}{
    \begin{tabular}{ c | c  }
     \toprule
     Meter & Percentage in corpus  \\ 
     \midrule
     4/4 & 26.65 \\
     2/4 & 22.03 \\
     3/4 & 20.44 \\
     6/8 & 13.00 \\
     3/8 & 5.22 \\
     other & 12.56 \\
     \bottomrule
    \end{tabular}}
\vspace{-1.75em}
\end{table}

\section{Model}
\label{sec:model}
The proposed pipeline is comprised of two main components:
i) a temporal prediction error model, in which our system predicts the next symbolic representation based on previous input symbols (similar to a language modeling task); ii) a peak detection algorithm that gets as input segmentation scores and outputs the segment boundaries. These modules are trained separately, and during inference, we cascade both modules to predict the final segmentation.

\subsection{Temporal Prediction Model}

Consider a single input files as $\vx = (x_1, x_2, \dots, \x_t)$, where $1 \leq t \leq T$. The length of $\vx$ varies between different inputs, hence $T$ is not fixed. Each $x_t$ is represented as four different input scalars such that $x_{t,1}, x_{t,2}, x_{t,3}, x_{t,4} \in \{0 \dots 128\}$, where these corresponds to different states of the note (128 represents the 128 possible unassigned midi values). Specifically, $x_{t,1}, x_{t,2}, x_{t,3}, x_{t,4}$ corresponds to: i) a note burst at the beginning of a bar; ii) a note burst occurring at the middle of the bar; iii) a note continuation at the beginning of a bar; iv) a note continuation at the middle of the bar, respectively. 


We follow the aforementioned input representation approach so the model could distinguish between the different states of a symbolic musical note (e.g., beginning of a bar, beginning of a note, etc.).  We sample each note at a constant sampling rate of size $\alpha$ to produce each $x_i$ (e.g., 8$^{th}$ notes, 16$^{th}$ notes, etc.).

We now turn in to describing our temporal prediction model. Given a training set of $n$ examples $\S = \{\vx_i\}_{i=1}^n$, our goal is to minimize the following objective function, 
\begin{equation}
\label{eq:obj}
\ell(\S) = \sum_{i=1}^n \sum_{t=1}^T \ell(x^i_t ; x^i_{0, \cdots t-1}),
\end{equation}
where $\ell$ is the negative log likelihood loss function. The proposed model is comprised of an embedding layer to transform input scalars to dense vectors. To model sequential dependencies we use a Recurrent Neural Network (RNN) (specifically an LSTM network using one hidden layer), and a linear projection layer followed by a softmax normalization. The model is optimized using back propagation through time.



\begin{figure}[t!]
    \vspace{-2.2em}
    \centering
    \includegraphics[width=1.0\linewidth]{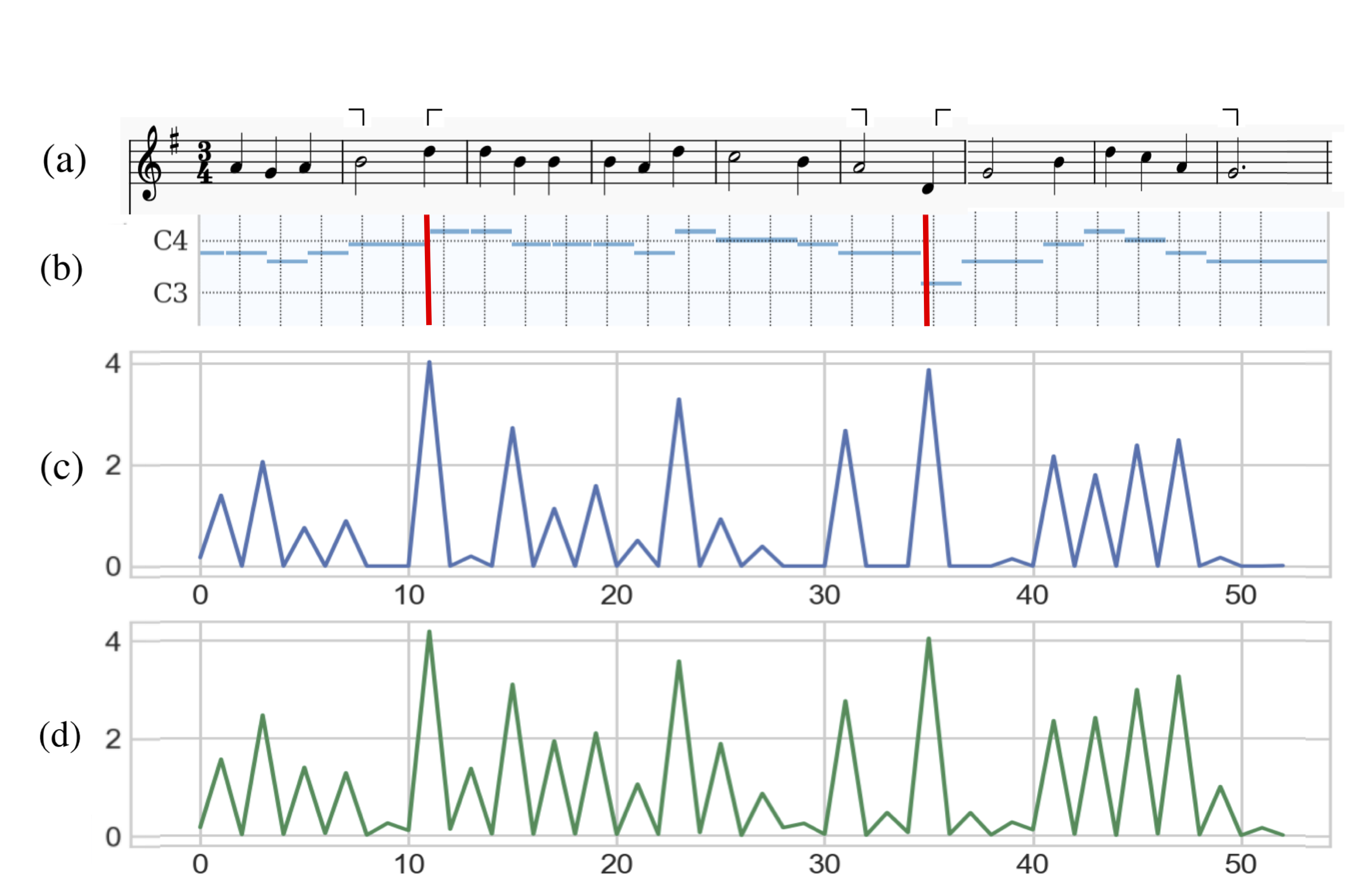}
    \caption{An example for a song in our test set. (a) the note format (opening and closing brackets mark beginning and end of human segments). (b) the midi format (human segments marked in red), (c) the corresponding normalized model temporal prediction error plot and (d) the corresponding original model temporal prediction error plot}
    \label{fig:loss_plot}
\vspace{-1.5em}
\end{figure}

\subsection{Peak detection}
During inference time, we calculate the negative log likelihood loss as in Eq.~\ref{eq:obj}. Specifically, we compute the loss at each time step $t$. Similarly to~\cite{michel2016blind} we ignore the first several time-steps as the system predicts probabilities conditioned by the preceding frames, hence cannot be expected to give meaningful results for the first inputs. Next, similarly to~\cite{adi2016automatic} we smooth the losses outputs by subtracting adjacent frames as follows,
\vspace{-1em}

\begin{equation}
\begin{split}
n\ell(j) = \max(0, a\cdot\ell(x^i_{j+1}; \cdot)+\ell(x^i_{j}; \cdot)-\\\ell(x^i_{j-1};\cdot)
-b\cdot\ell(x^i_{j-2}; \cdot))
\end{split}
\end{equation}

where $n\ell(j)$ is the normalized loss in the $j$'th place of the input, and $a$ and $b$ are two more hyper-parameters added to stabilize the peak prediction procedure. We perform the peak prediction on top of the normalized losses, where we control the level of segmentation by shifting the prominence of the peak detection. See Figure~\ref{fig:loss_plot} for an example prediction of the proposed method. We specifically visualize the musical notes, midi-format, loss function, and normalized loss function together with the boundaries of the phrases. 

The peak detection procedure is done by controlling the prominence $\delta$. To find the prominence for the peak detection of a certain song we perform a binary search (increasing the prominence for more peaks and decreasing it for fewer peaks) where we limit the number of peaks to be less than a certain threshold. We set the threshold to be the number of bars in the song divided by 2. In other words, the maximum amount of peaks that we allow the model to segment is the number of bars divided by 2 (notice the model can segment less than that).

\noindent {\bf Constraint Search.} To further improve model performance, we constraint the search space of the boundaries detection. We limit the search using two main constraints: i) \emph{Pause} selects boundaries after musical rests and doesn't select peaks within a range of distance $\Delta$ from each rest; ii) \emph{Bar} selects loss peaks only if they are at the beginning of a bar. In the case of songs that do not start at the beginning of a bar (denoted as anacrusis), we shift the bar notation such that the first strong beat occurs on the first onset.

\subsection{Ensemble learning}
We follow the ensemble approach by training $k$ different models for each time-signature independently. To construct the final outputs segmentation, we perform a union vote phase where we take all the segment candidates from all models. Next, we sort the segments according to the number of votes they got from the different models. Lastly, we output the top segments until either we take all segments or we reach the pre-defined threshold (number of bars in the input divided by 2). 

\vspace{-0.2em}
\section{Experiments}
\label{sec:exps}

\subsection{Dataset}
We evaluate our model on the Essen Folksong Collection~\cite{schaffrath1995essen} dataset, a widely used benchmark for evaluating boundary detection predictions of musical phrases. It consists of 6,236 European folksongs in a symbolic format, most of them from Germany. The corpus was annotated by musical experts, containing phrases boundaries at the note level. We used the standard train/test split, where we randomly sampled 10\% of the training set for validation. We consider six different musical time-signature groups. Table~\ref{tab:db_stats} summarizes the distribution over the different time signatures in the Essen collection.

\begin{table}[t!]
\centering
\caption{Comparison of our model to other baseline models on the Essen Folksong dataset. Results are listed in increasing order by F-score per setting. Since some of the models are rule-based they were evaluated on the entire data-set and thus denoted by: *}
\label{tab:main}
\resizebox{\columnwidth}{!}{%
\begin{tabular}{ l | l | c | c | c | c}
 \toprule
  & Model & Precision & Recall & F-Score & R-Value  \\ 
 \hline
 \multirow{8}{*}{\rotatebox[origin=c]{90}{Rule based}} 
  & $\Delta$IOI* \cite{cenkerova2018crossing} & 79 & 54 & 58 & 67 \\
  & Pause* \cite{cenkerova2018crossing} & 98 & 48 & 60 & 63 \\
  & Meter* \cite{temperley2004cognition} & 59 & 70 & 61 & 65 \\
  & Meter Finder* \cite{toiviainen2006autocorrelation} & 70 & 64 & 64 & 72 \\
  & LBDM* \cite{cambouropoulos2001local} & 81 & 60 & 65 & 71 \\
  & Grouper without meter* \cite{temperley2004cognition} & 68 & 66 & 66 & 68 \\
  & $\Delta$IOI, Meter Finder, Pause* \cite{cenkerova2018crossing} & 64 & 81 & 68 & 63 \\
  & Grouper with meter* \cite{temperley2004cognition} & 77 & 73 & 74 & 78 \\
 \hline
 \multirow{5}{*}{\rotatebox[origin=c]{90}{Unsup.}}
  & IdyOM \cite{pearce2010melodic} & 76 & 50 & 58 & 64\\
  & RBM (10-gram) \cite{lattner2015probabilistic} & 83 & 50 & 60 & 64\\
  & Hybrid IdyOM \cite{pearce2008comparison} & 87 & 56 & 66 & 69\\
  & $\Delta$IOI Compound (BIC) \cite{cenkerova2018crossing} & 92 & 68 & 75 & 77 \\
  & \textbf{Ours} & 77 & \textbf{81} & \textbf{77} & \textbf{82} \\
 \hline
 \multirow{5}{*}{\rotatebox[origin=c]{90}{Sup.}} 
  & Ripper \cite{van2020rule}  & 78 & 63 & 69 & 73\\
  & RandomForest \cite{van2020rule}  & 83 & 69 & 76 & 77\\
  & DOP-Markov \cite{bod2002memory}  & 77 & 86 & 81 & 81\\
  & CNN-CRF \cite{zhang2020symbolic}    & - & - & 82 & - \\
  & BI-LSTM-CRF \cite{zhang2020symbolic} & - & - & 84 & - \\
 \bottomrule
\end{tabular}}
\end{table}

\subsection{Evaluation Function}
Following previous work on music and phoneme segmentation~\cite{cenkerova2018crossing, bod2002memory, kreuk2020self, michel2016blind} we evaluate the performance of the proposed method and the evaluated baselines using Precision (P), Recall (R), and F1-Score. Previous studies~\cite{rasanen2009improved}, pointed out a potential drawback in the F1-Score for boundary detection due to its sensitivity to over-segmentation. A naive segmentation model that outputs a boundary every constant time of seconds (in our case notes) may yield a relatively high F1-Score by achieving high recall at the cost of low precision. Thus the authors of~\cite{rasanen2009improved} suggested using the R-Value metric defined as follows:
\begin{align*}
\text{R-value} = 1 - |r_{1}| - |r_{2}|,
\end{align*}
\begin{equation}
r_{1}= \sqrt{(1-R)^2+(OS)^2}, r_{2}= \frac{(-OS+R-1)}{\sqrt{2}} 
\end{equation}
where R is the Recall value, P is the Precision value and OS is the over segmentation measure defined as $OS=R/P-1$. For a fair comparison we calculate the R-values based on the mean precision and recall in all settings.  

\subsection{Experimental Setup}
We train $k$ different models as part of the ensemble. We launch a grid search over learning rate, batch size, and input sequence length and used the best configuration over the held-out set. Models were trained with one of the following input pre-processing pipelines: transposing each song by 5 semitones or scaling all songs to the same musical key. These are standard pre-processing pipelines in symbolic music modeling~\cite{simon2017performance, chu2016song}. We train the models with the ADAM optimizer, with one hidden layer and a constant dropout rate of 0.2. Finally, we perform a peak detection procedure, as described in section~\ref{sec:model} to each model separately and aggregate the results for our final model.

\subsection{Results}
We compare our method to unsupervised baselines proposed by~\cite{cenkerova2018crossing, lattner2015probabilistic, pearce2008comparison}, rule-based methods proposed by~\cite{cenkerova2018crossing, temperley2004cognition, toiviainen2006autocorrelation, cambouropoulos2001local} and supervised models as in~\cite{zhang2020symbolic, van2020rule, bod2002memory}. We additionally compare to a naive approach denoted as Pause. It was shown to be a comparative baseline in previous work~\cite{pearce2010melodic}. The basic idea is straight-froward, where we annotate a  segment boundary on every rest in the music melody. 

\begin{table}[t!]
	\centering
	\caption{The proposed model results based on the different time signatures in the test set}
	\label{tab:analysis}
	\scalebox{0.8}{
    \begin{tabular}{ c | c | c | c | c }
     \toprule
     Meter & Precision & Recall & F-Score & R-Value  \\
     \midrule
     4/4 & \textbf{81.36} & \textbf{84.08} & \textbf{81.70} & \textbf{85.06} \\
     2/4 & 75.30 & 76.36 & 74.14 & 79.30 \\
     3/4 & 71.33 & 81.33 & 73.67 & 76.77 \\
     6/8 & \textbf{83.51} & \textbf{87.02}& \textbf{84.69} & \textbf{87.11} \\
     3/8 & 61.12 & 83.16 & 68.60 & 61.43 \\
     other & 80.92 & 72.76 & 73.94 & 79.41 \\
     \bottomrule
    \end{tabular}}
\vspace{-1.2em}
\end{table}

Results are summarized in Table~\ref{tab:main}. Results suggest the proposed method reaches state-of-the-art results considering both  F-Score and R-Value. When considering recall our model is comparable to the method proposed in~\cite{cenkerova2018crossing} however with a significantly better precision value. Interestingly, although the proposed method is inferior to the supervised setting, it significantly closes the gap between supervised and unsupervised methods for symbolic music segmentation, which leaves room for future work to be done.

To better understand how the proposed method performs under different time-signatures, we report Precision, Recall, F-Score, and R-value as a function of six different time signatures in Table~\ref{tab:analysis}. It can be seen that the proposed model performs significantly better on 4/4 and 6/8 time signatures. The lowest results are for 3/8 meter which is the least common of the above 5 time signatures in the Essen Folksong Dataset.

\subsection{Ablation}
In order to further assess the importance of each of the components composing our overall method, we analyze a few different model variations which contain only a subset of the modules comprising our pipeline. Specifically, we define ``Single-Temp'', a single model trained with using temporal prediction errors additionally with the search constraints that were described in section~\ref{sec:model}. The second method, ``Multi-Temp'', uses a separate model for each time signature. The last one denoted as ``Ensemble-Multi-Temp'' is the full proposed method, aggregating the results of a few different models for each time signature separately. A new hyper-parameter search was performed for each ablation configuration. Results are summarized in Table~\ref{tab:ensamble}.


It can be seen that splitting the learning procedure into a few different models improves the results, especially with respect to the precision (Single-Temp reaches 75.62 precision while Multi-Temp reaches  80.51 precision). This brings the model to an overall high precision on account of low recall. Whereas for the proposed model (Ensemble-Multi-Temp), aggregating the results of a few ensemble models balances the precision and recall yielding a total higher F-Score and R-Value. 

Next, we present the results of the proposed method (for simplicity using a single model only) with and without limiting the chosen boundaries. In other words, we evaluate the contribution of the constraint search described in Section~\ref{sec:model} to the final model performance. We do this for both the "Pause" and "Bar" search constraints. Results are summarized in Table~\ref{tab:temp_bars}.

\begin{table}[t!]
\vspace{-0.4em}
	\small
	\centering
	\caption{Model performance using single, multiple temporal prediction models and an ensemble}
    \label{tab:ensamble}
	\scalebox{0.8}{
    \begin{tabular}{ c | c | c | c | c }
     \toprule
     Model & Precision & Recall & F-Score & R-Value  \\ 
     \midrule
     Single-Temp & 75.62 & 74.32 & 73.14 & 78.66 \\
     Multi-Temp & 80.51 & 75.59 & 75.96 & 80.95 \\
     Ensemble-Multi-Temp & \textbf{76.96} & \textbf{80.81} & \textbf{77.03} & \textbf{81.53} \\
     \bottomrule
    \end{tabular}}
\end{table}

\begin{table}[t!]
\vspace{-0.3em}
	\small
	\centering
	\caption{The single-temp model without the use of the pause and bar search constraints, within comparison to the bar and pause search constraints without temporal prediction}
    \label{tab:temp_bars}
	\scalebox{0.8}{
    \begin{tabular}{ c | c | c | c | c }
     \toprule
     Model & Precision & Recall & F-Score & R-Value  \\ 
     \midrule
     Bars,Pause & 47.34 & 89.83 & 60.87 & 19.06 \\
     Bars & 45.47 & 83.32 & 57.63 & 22.2 \\
     Pause & 98.28 & 48.74 & 61.09 & 63.76 \\
     \midrule
     Single-Temp & \textbf{75.62} & \textbf{74.32} & \textbf{73.14} & \textbf{78.66} \\
     Single-Temp without Pause & 64.16 & 61.82 & 60.97 & 68.61 \\
     Single-Temp without Bars & 59.7 & 63.67 & 60.5 & 66.34 \\
     Single-Temp without Bars,Pause & 46.86 & 52.07 & 48.39 & 54.53 \\
     \bottomrule
    \end{tabular}}
\vspace{-1.9em}
\end{table}

Using a temporal prediction error model without constraining the search yields considerably low results on the evaluated benchmark with regards to F-Score. Adding the Pause and Bar methods to the model provides a significant increase of the F-Score values, from 48.39  to 73.14, which is already close to the previous best state-of-the-art unsupervised setting by itself.

\begin{figure}[t!]
\vspace{-2em}
    \centering
        \scalebox{0.78}{
        \includegraphics[width=1.0\linewidth]{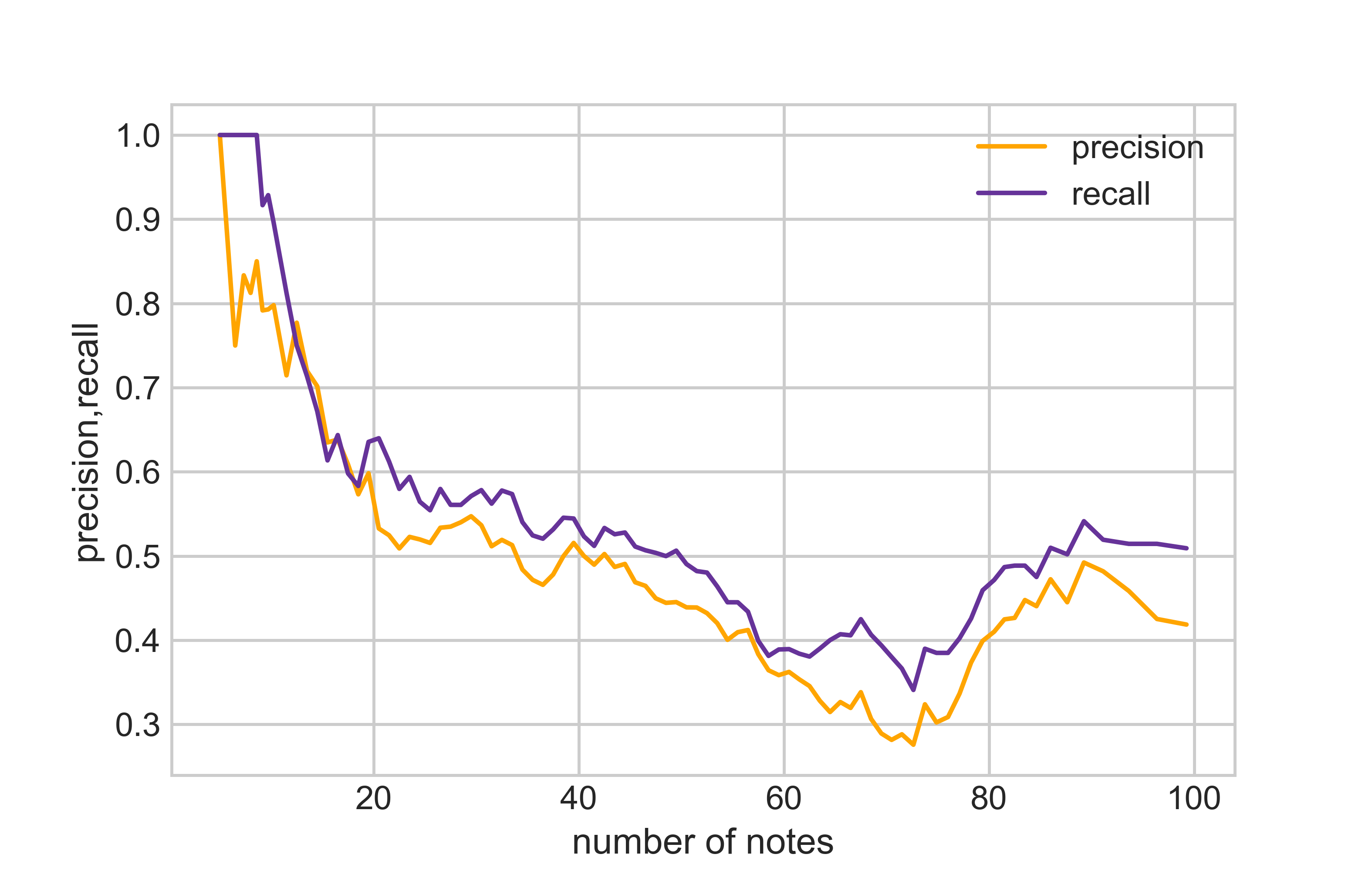}}
    \caption{Precision and Recall as a function of the number of notes for the Single-Temp model without the search constraints.}
    \label{fig:bar}
\end{figure}

Next, we plot the precision and recall as a function of the number of notes in a given input example for a single model without the search constraints. This obviously excludes the first few notes that were discarded for evaluation as mentioned in section~\ref{sec:model}. Figure~\ref{fig:bar} depicts the results. Increasing the number of notes in a song decreases the quality of the overall segmentation. We hypothesize this is due to the RNN model having trouble capturing long-term dependencies. Interestingly, we observed an increase in performance after 80 notes, however, due to a low number of testing files in that region it is hard to make a firm conclusion.

\begin{table}[t!]
\vspace{-0.1em}
	\centering
	\caption{The different model F scores by Time signature}
    \label{tab:time_s}
	\scalebox{0.75}{
    \begin{tabular}{ c | c | c | c }
     \toprule
     Meter & Single-Temp without Bar,Pause & Single-Temp & Ensemble-Multi-Temp  \\
     \midrule
     4/4 & \textbf{49.88} & \textbf{79.73} & \textbf{81.70}  \\
     2/4 & 51.65 & 73.87 & 74.14  \\
     3/4 & 48.37 & 69.19 & 73.67  \\
     6/8 & \textbf{50.02} & \textbf{78.57} & \textbf{84.69}  \\
     3/8 & 47.33 & 61.37 & 68.60 \\
     other & 39.36 & 64.72 & 73.94\\
     \bottomrule
    \end{tabular}}
\vspace{-1.75em}
\end{table}

Lastly, we report the results of a single model with and without the constraint search compared to the full pipeline for different time-signatures separately on Table~\ref{tab:time_s}. Results suggest that the most common time signatures in the data-set have a similar F-Score when considering the temporal prediction model only, whereas the less common time signatures such as 3/8 or other less common meters have a lower F-score. Nevertheless, the addition of the Pause and Bar constraints greatly affects the results providing a significantly higher F-Score for both 4/4 and 6/8. This is an indicator that the performance differences between different time-signatures, considering the final proposed model, are due to the constraint search being biased towards specific time signatures. This also has to do more with the data-set bias and less with the temporal prediction method. 
\vspace{-0.4em}
\section{Conclusion $\&$ Future Work}

We empirically demonstrated the efficiency of using temporal prediction errors for symbolic music segmentation and showed how to enhance this procedure by using an ensemble method. Our model reached SOTA results on the Essen Folksong Dataset when considering F1-score, and R-Value, under the unsupervised setting. This is a promising result in terms of closing the gap between unsupervised and supervised methods for symbolic music segmentation.

For future work, we would like to explore the semi-supervised setting, where we provide the model with a limited amount of human-labeled segments. Additionally, we would like to explore extending this work to both performance-midi segmentation as well as music audio segmentation.\\

\textbf{Acknowledgments:} This research has been partly funded by Israel Science Foundation grant 1340/18.
\label{sec:con}

\bibliographystyle{IEEEtran}
\bibliography{bib}
\end{document}